\documentstyle[epsfig,lens]{mn}

\title[Magnetic Lensing]{Magnetic Lensing near Ultramagnetized Neutron Stars} 

\author[N. J. Shaviv, J. S. Heyl \& Y. Lithwick]
{Nir J. Shaviv, Jeremy S. Heyl and Yoram Lithwick \\ 
Theoretical Astrophysics 130-33, California Institute of Technology, \\
Pasadena, CA 91125}
\date{Submitted to MNRAS, October 29}

\pagerange{1--12}
\pubyear{1998}

\begin{document}

\maketitle

\label{firstpage}
\begin{abstract}

Extremely strong magnetic fields change the vacuum index of
refraction. This induces a lensing effect that is not unlike the
lensing phenomenon in strong gravitational fields. The main difference
between the two is the polarization dependency of the magnetic
lensing, a behaviour that induces a handful of interesting effects. The
main prediction is that the thermal emission of neutron stars with
extremely strong magnetic fields is polarized - up to a few percent
for the largest fields known. This potentially allows a direct method
for measuring their magnetic fields.

\end{abstract}



\section{Introduction}
\label{sec:intro}

The most intense magnetic fields in the universe are found near
neutron stars.  Soon after the discovery of radio pulsars
(\cite{Hewi68}), \jcite{Gold68} argued that the steady increase in
their periods could be explained if they were strongly magnetized
neutron stars with dipolar fields of $\sim 10^{12}$~G.  Since that
time, the variety of radio pulsars has expanded.  These neutron stars
have magnetic fields ranging from $10^8$~G for the millisecond pulsars
to $2.1 \times 10^{13}$~G for PSR 0157+6212 (\cite{Arzo94}), the most
strongly magnetized radio pulsar discovered to date.

A dozen years after the discovery of radio pulsars, the increased
scrutiny of the sky at $\gamma$-ray energies uncovered a new class of
objects, the soft-gamma repeaters (SGRs).  In the most successful model
for these energetic sources, the crust of a strongly magnetized ($B
\sim 10^{14-15}$~G) neutron star or ``magnetar'' periodically fractures,
releasing a large amount of energy into the magnetosphere which is
observed as soft-gamma radiation (\cite{Thom95}).  Until the past
year, the evidence for the intense magnetic fields surrounding SGRs
was circumstantial.  As \jcite{Gold68} argued for radio pulsars,
\jcite{Kouv98} precisely measured the spin down of SGR~1806-20 and
inferred a dipole field of $8 \times 10^{14}$~G.  \jcite{Kouv981900}
measured ${\dot P}$ of SGR~1900+14 and find it to be consistent with 
$B = 2 - 8 \times 10^{14}$~G.

In the past decade, more sensitive X-ray telescopes have also
discovered a new class of sources which may have extremely strong
magnetic fields.  The anomalous X-ray pulsars (AXP) are soft X-ray
sources and they are often associated with supernova remnants.  Unlike
the majority of X-ray pulsars, these objects have periods of several
seconds which steadily and quickly increases.  Furthermore, no binary
companions have been found near these objects.  Although no single
picture has yet emerged to explain AXPs, a strong candidate model is
that a neutron star cooling through a strongly magnetized envelope
(\cite{Heyl97kes}) or the decay of a neutron star's strong magnetic
field (\cite{Heyl98decay}; \cite{Thom96}) powers the X-ray emission
from these objects.  Furthermore, such a strong magnetic field ($B
\sim 10^{15}$~G) can also account for the observed spin-down rate of
these objects (\eg \cite{Vasi97b}).

Intense magnetic fields of $B \gtrsim 10^{12}$~G strongly affect
physical process in and near neutron stars -- ranging from cooling
(\cite{Heyl97magnetar}; \cite{Shib95}), to atmospheric emission
(\cite{Raja97}; \cite{Pavl94}) and to the insulation of the core
(\cite{Heyl98numens}; \cite{Heyl97analns}; \cite{Scha90a};
\cite{Hern85}).  Yet stronger fields such as those associated with AXPs
and SGRs alter the propagation of light through the magnetosphere by
way of quantum-electrodynamic (QED) processes and may further process
the emergent radiation (\cite{Heyl97hesplit}; \cite{Heyl97index};
\cite{Bari95a}; \cite{Bari95c}; \cite{Adle71}).

Not only will QED alter the spectra from these objects, it also
creates a lens surrounding the neutron star which magnifies and
reimages the surface.  In a sufficiently strong field, the index of
refraction for photons with their magnetic fields directed
perpendicularly to the local magnetic field can be significantly
larger than unity (\cite{Heyl97index}).  Furthermore, it is a strong
function of both the strength of the magnetic field and its direction.  In
this paper, we will examine the effects of magnetic lensing on the
observations of strongly magnetized neutron stars.

\section{Index of refraction}
\label{sec:index}

In the presence of a strong external field, the vacuum reacts,
becoming magnetized and polarized.  The index of refraction, magnetic
permeability, and dielectric constant of the vacuum are
straightforward to calculate using quantum electrodynamic one-loop
corrections (\cite{Klei64a}, \cite{Klei64b}, \cite{Erbe66},
\cite{Adle71}, \cite{Bere82}, \cite{Miel88}).  \jcite{Heyl97index}
calculate the index of refraction as a function of field strength to
the one loop order.  The magnetic field will strongly bend light as it 
propagates only if $n-1 \sim 1$.  

For $B \gg\ B_\rmscr{QED} \approx 4.4 \times 10^{13}$~G, the index of
refraction becomes significantly larger than unity if the wave has its
magnetic field polarized perpendicular to the local magnetic field
($\perp$-mode)\footnote{Note that the convention used here is
consistent with \jcite{Adle71}, \jcite{Bere82} (\S 130) and
\jcite{Heyl97index} in which the $\perp$ and $\parallel$ modes
describe the polarization direction of the ray's {\em magnetic} field
relative to the local magnetic field. \jcite{Bari95a} and other
authors use the wave's {\em electric} field to denote the
polarization.}.  To one loop, the waves with their magnetic field
directed in the plane containing the local magnetic field, ${\bf B}$,
and the wavenumber, ${\bf k}$ ($\parallel$-mode), have $n_\parallel -
1 < \alpha / (6 \pi)$ regardless of the strength of the field
($\alpha$, the fine-structure constant is approximately $1/137.04$).
This second mode is essentially unaffected by lensing.

In fields where $n_\perp - 1 \sim 1$, \jcite{Heyl97index} find that 
$n_\perp - 1 \propto B$ with sufficient precision.  Specifically, for $B
> B_\rmscr{QED}/2$
\def\cond{{\hskip 0pt}}
\figcomment{\def\cond{{\hskip -4pt}}}
\ba
n_\perp \cond \hskip -4pt &=& \hskip -3pt
\cond 1 + \frac{\alpha}{4\pi} \sin^2\theta \Biggr [ \frac{2}{3} \xi
- \left ( 8 \ln A - \frac{1}{3} - \frac{2}{3} \gamma \right ) 
 \nonumber \\
&& \cond \cond 
 - \left ( \ln\pi + \frac{1}{18} \pi^2 - 2 - \ln \xi \right ) \xi^{-1}
- \left ( -\frac{1}{2} - \frac{1}{6} \zeta(3) \right ) \xi^{-2} 
\nonumber \\ 
& & \cond \cond 
  - \sum_{j=3}^\infty \frac{(-1)^{j-1}}{2^{j-2}} 
 \left ( \frac{j-2}{j(j-1)} \zeta(j-1) + \frac{1}{6}
\zeta(j+1) \right ) \xi^{-j} \Biggr ] 
\figcomment{\nonumber \\
& & \cond \cond }
 + {\cal O}\left[ \left (\frac{\alpha}{2\pi} \right )^2 \right ], 
\ea
where $\xi = B/B_\rmscr{QED}$ and $\theta$ is
the angle between ${\bf B}$ and ${\bf k}$. $\gamma$ is Euler's
constant and $\ln A$ is related to the derivative of the zeta
function: $\ln A\equiv 1/12-\zeta^{(1)}(-1)=0.2487...$~.  The linear
approximation is accurate to 20~\% for the value of $n-1$ for fields
stronger than $3.9 \times 10^{14}$~G where $n - 1 = 2.9
\times 10^{-3}$.  For this same field strength, $n-1$ for the parallel
polarization is ten times smaller.  For stronger fields this ratio
increases.  Thus, we set $n_\parallel=1$.

The function $f(B) \equiv (n_\perp - 1)/\sin^2 \theta$ contains the portion
of the index of refraction which depends on the strength of the
external field.  The linear approximation is given by $f(B) =
\alpha / (6 \pi) \times  B/B_\rmscr{QED}\equiv B/B_0$.  The second loop
corrections to the index of refraction are expected to be smaller by a
factor of $\alpha$ regardless of the strength of the field
(\cite{Ditt85}). 

\section{$\perp$-Ray propagation}
\label{sec:rayprop}

In the limit of geometric optics, light rays obey Hamilton's equations
with $H = \omega = c k/n({\bf k}, {\bf x})$ (\cite{Land2}, \S 53).
Since the index of refraction depends both on position and wavevector,
the resulting equations are nontrivial,
\be
{\dot {\bf x}} = \nabla_k \omega ~\rmmat{and}~ {\dot {\bf k}} =
-\nabla_x \omega.
\label{eq:ray_traj}
\ee
Each of these gradients may be calculated 
for the $\perp$-mode using the 
index of refraction,
$n_\perp=1+[1-({\bf k\cdot B}/kB)^2]f(B)$:
\ba
\nabla_{\!x} \omega  &=& 
{ckf\over n_\perp^2}
\Bigg[
-\Big({d\ln f\over d\ln B}\sin^2\theta +2\cos^2\theta\Big)
 {{\bf \nabla}B\over B}
\figcomment{\nonumber \\
& & \hfill } 
+2\cos\theta{ {\bf \nabla}({\bf k\cdot B})\over kB}
\Bigg] \\
\nabla_{\!k} \omega &=& 
{c\over n_\perp}\frac{{\bf k}}{k} 
+{2cf\over n_\perp^2}\cos\theta
\Big({{\bf B}\over B}-\cos\theta
     {{\bf k}\over k}
\Big) \ .
\ea
As outlined in the previous section, the 
propagation depends on the 
polarization of the photon; thus, it is important to
understand how the photon's polarization evolves as the direction of
the field changes.

\subsection{Polarization evolution}
\label{sec:polar}

As the waves travel through the magnetosphere, they stay in the same
polarization mode (parallel or perpendicular) as long as
\be
\frac{n_\|}{n_\perp} \frac{1+n_\perp^2}{\Delta k} \frac{1}{r_n} \ll 1
\ee
where $r_n$ is the distance over which the magnetic field rotates by
one radian and $\Delta k$ is the difference between the wavenumbers of
the two polarizations.  Evidently, the adiabatic approximation holds
for the vicinity of the systems of interest (see \cite{Heyl98polar}
for further details).

A photon's polarization will decouple from the magnetically induced
polarization modes at a distance from the star which depends on the
strength of the field at the stellar surface and the energy of the
photon.  Higher frequency photons decouple later; therefore, the
observed polarization direction will depend on photon energy with
higher frequency photons tracing the magnetic moment of the star
projected onto the sky from later in the star's rotation.

Generally, the wavelength of the photons is much shorter than the
length over which the field geometry changes, so the decoupling occurs
where the magnetic field is much less than the QED critical value.  In
this limit, the wave decouples from the field after traveling for a
time,
\be
t_\rmscr{decouple}  \approx   4.6\hskip 2pt \rmmat{ms} \hskip 2pt
	\cond \left ( \frac{E}{1~\rmmat{keV}} \right )^{1/5}
	\cond \left ( \frac{B_p \sin\theta}{10^{14}~\rmmat{G}} \right )^{2/5}
	\cond \left ( \frac{R}{10~\rmmat{km}} \right )^{6/5} \cond,
\label{eq:decouple}
\ee
where $B_p$ is the strength of the magnetic field at the surface, $R$
is the radius of the neutron star and $E$ is the energy of the
photon.

The typical value of $t_\rmscr{decouple}$ for high $B$ neutron stars
is such that decoupling will take place well within the light cylinder
yet far enough from the stellar surface such that the local magnetic
field where the photons decouple is aligned with the apparent magnetic
axis, irrespective from where the photons originated.  This simplifies
the calculation of the two polarization images -- The two principle
polarization directions at the observer are the $\parallel$ and
$\perp$ modes even though the polarization directions from where the
photons orginiated were different.

\def\usrf{\hat{{\bf s}}}
\def\xd{\hat{{\bf x}}}
\def\yd{\hat{{\bf y}}}
\def\zd{\hat{{\bf z}}}
\def\vg{{\bf v}_\rmscr{g}}
\def\kv{{\bf k}}
\def\asym{{\cal A}}
\def\deg{^\circ}
\def\para{\parallel}

\section{The appearance of a magnetar}

The nontrivial index of refraction depends on the direction and
magnitude of the local magnetic field. Light rays will therefore
travel in bent trajectories and the optical appearance of the
magnetized neutron star will therefore not be trivial. From equivalent
optical configurations (e.g. a black body embedded within a
nonspherical medium of index $n$), it is clear that the images of the
star will be distorted and can have a varying apparent surface area,
even though it is spherical.

We use the ray tracing algorithm to find the image of a magnetar
having a dipole magnetic field. For each observing direction
characterized by an inclination angle above the magnetic equator, a
different image may result. These images can be compared to the
unmagnetized case which adequately describes the $\parallel$
polarization. Each image can then be averaged in order to compose a
light curve. And finally, all the images can be properly averaged
together, to get the total emission properties of the NS.

\subsection{The ray tracing algorithm}

Using the Hamiltonian formalism and the related adiabatic and
geometrical optics approximations (which hold extremely well in our
limit), we know that each polarization mode (either the $\para$ or
$\perp$ modes) will evolve {\em separately} according to eq.~\ref{eq:ray_traj}:
\begin{equation}
\dot {\bf x}_{\perp,\para} = \nabla_k \omega_{\perp,\para}~~\rmmat{and}~~
\dot {\bf k}_{\perp,\para} = -\nabla_x \omega_{\perp,\para}.
\end{equation}
Since we are interested in the regime in which the index of refraction
$n_\para$ of the $\para$-mode can be assumed to be 1, its evolution becomes
trivial. The image observed in the $\perp$ polarization will not
however be straight forward and its light ray trajectories will be
bent.

To construct the image observed by a distant observer, we can use the
time reversal symmetry of the light trajectory and trace the light
rays from the observer to the NS and not vice versa. We place a screen
at a large distance from the star and divide it into pixels at the
desired resolution (e.g.~$300^2$). From each pixel, we trace back the
path of a light ray leaving in a direction perpendicular to the
screen. The ray is then followed until it intersects the NS surface or
until it is found to have an increasing radial component, implying
that it has missed the star. The configuration is depicted in figure
\ref{ray_tracing_fig}.

\def\figray{
\begin{figure}
\figcomment{\centerline{\psfig{file=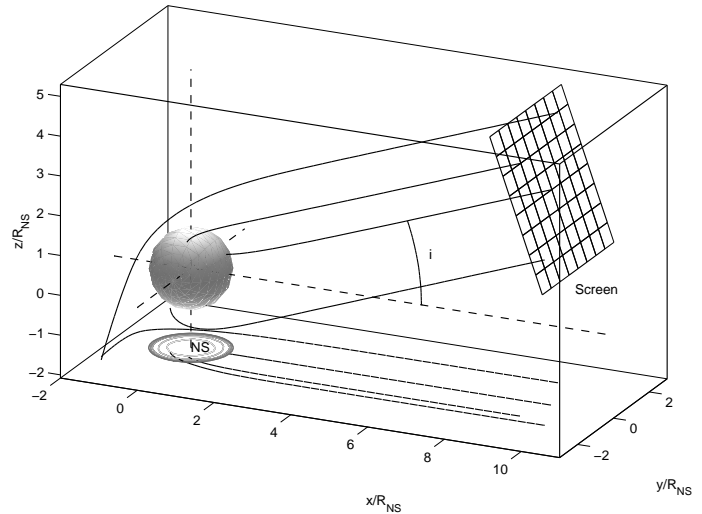,width=\figurewidth}}}
 \caption{The ray tracing algorithm. Ray trajectories are followed
 from a distant screen towards the star using the Hamiltonian
 equations of motion. The rays can then either intersect the star or
 miss it. Since the magnetic field is assumed to be that of a dipole
 (aligned with the $z$-axis), it has axial symmetry and different
 images will be distinguishable by the observer's magnetic inclination
 angle $i$.}
\label{ray_tracing_fig}
\end{figure}}
\figcomment{\figray}

It is clear from Kirchhoff's law that if we look at a light ray
trajectory that intersects the surface, then if the observer and the
object are in thermal equilibrium, we will find that the specific
intensity $I$ (flux per unit sterad) of the ray going from the
observer to the object should be equal to that of the opposite ray
going from the star. Since the observer is at a large distance and has
$n\rightarrow 1$ in its vicinity, the specific intensity is just
$I=B(T)=\sigma T^4 / \pi$, where $B(T)$ is the black-body source
function (once the ray is in an $n=1$ region).  Hence, if the emission
from the NS is that of a black-body, then the rays coming from each
element will simply carry a specific intensity of $B(T)$ where $T$ is
the local effective photospheric temperature at the point where the
ray intersects the surface. If the NS has a uniform temperature then
the flux observed by an observer at a given direction is simply
proportional to the solid angle subtended by the object. As we shall
see, this apparent solid angle will be larger from all $4 \pi$
possible observer directions implying that the total luminosity
emitted at a given temperature is larger in the presence of a magnetic
field. This does not necessarily imply that a magnetized NS will
shine brighter. As we shall see in \S5, the changed surface
properties will reduce the temperature to compensate for the higher
emission efficiency, leaving an unchanged luminosity.

\subsection{Image properties}

A few resulting images can be seen in figure \ref{sample_images} which
depicts the magnetic latitude and longitude observed at several
inclinations for both a weak field and a strong one. We find that in
the limit of a weak field (one for which the correction to the index
of refraction is less than half), the effects of the magnetic lensing
are to slightly distort the image through the imaging of otherwise
inaccessible surfaces (surfaces that do not have a direct line of
sight) and to have the images appear to subtend a larger surface area.
In strong fields with corrections larger than a half to the index of
refraction, on the other hand, the images are distorted enough as to
have a topologically nontrivial image form in which some surfaces are
seen twice (or more) with changing parities.

\def\figsix{
\begin{figure*}
 \figcomment{\centerline{\psfig{file=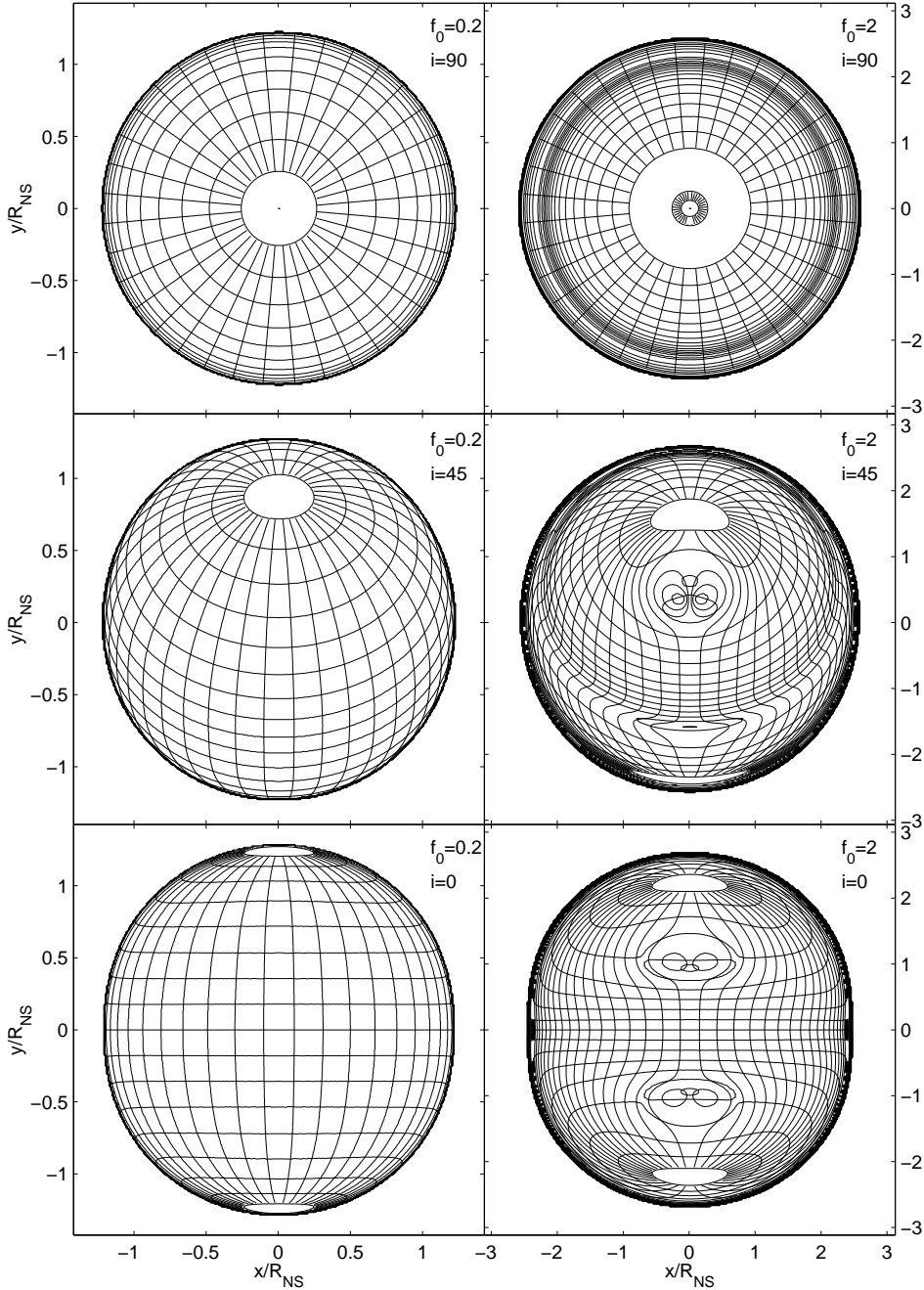,width=5.0in}}}
 \caption{The magnetic latitude and longitude of the surface of the NS
 as seen with the $\perp$ polarization by a distant observer at
 different inclinations above the magnetic equator (with spacing of
 $10\deg$ in each direction). The equatorial magnetic field strengths
 of $2.3\times10^{16}$~G and $2.3\times10^{17}$~G corresponding to
 $f_0=0.2$ and $f_0=2.0$. An observer of a ``weak'' field NS will see
 a larger image and more than $2 \pi$ sterad of the surface. If the NS
 is a black-body, more light will be seen with the $\perp$
 polarization than with the $\para$ polarization (for which the image
 is unchanged).  An observer of a highly magnetized NS will see a more
 complex image in which the same surface element can appear twice (or
 even more).}
\label{sample_images}
\end{figure*}}
\figcomment{\figsix}

The critical equatorial field needed to trigger the nontrivial
behaviour is $2.8 \times 10^{16}$~G for which $f_0\equiv f(B_\rmscr{equator})
= 0.25$ on the equator and $f(B_\rmscr{pole})=0.5$ on the poles. At the
latter field strength (for which $n_\perp$ can reach $1.5$), the group
velocity is no longer a monotonic function of $\kv$, allowing multiple
light ray solutions with the same $\vg$ (3 instead of 1).

If the NS has a uniform surface temperature, its luminosity will just
be proportional to the surface area observed. This apparent surface area as a
function of the magnetic inclination $i$ of the observer can be well fitted
with the following functional form:
\begin{equation}
A' = a_0 + \sum_{n=1}^{4} a_n \cos^n(2 i).
\label{fitting_eq}
\end{equation}

One can also define an asymmetry parameter as the normalized
difference between the area observed above the equator and that
observed above the pole:
\begin{equation}
 \asym \equiv  {A'(i=0\deg) - A'(i=90\deg) \over A'(i=0\deg) +
 A'(i=90\deg)} = {( a_1 +  a_3) \over a_0}
\end{equation}
The value of the fit parameters and the total asymmetry are shown in
figure \ref{asym_fit_fig}. Evidently, one finds
simple expressions for the fitting constants in the limit of weak fields:
\begin{eqnarray}
\label{eq:expA}
 \asym \cond \cond  & \approx & \cond \cond 0.095 f_0 = { B_0 \over
 1.20\times 10^{18} G } ~~\rmmat{for}~~B_0 \lesssim 10^{16}
 G~~\rmmat{and:} \\  a_1 \cond \cond & \approx & \cond \cond 0.89 \asym, ~a_2
 \approx -0.08 \asym, ~a_3 \approx 0.11 \asym, ~a_4 \approx -0.04
 \asym. \nonumber 
\end{eqnarray}
With this fit, we can construct a light curve for a thermally uniform
rotating NS.

\def\figfits{
\begin{figure}
\figcomment{\centerline{\psfig{file=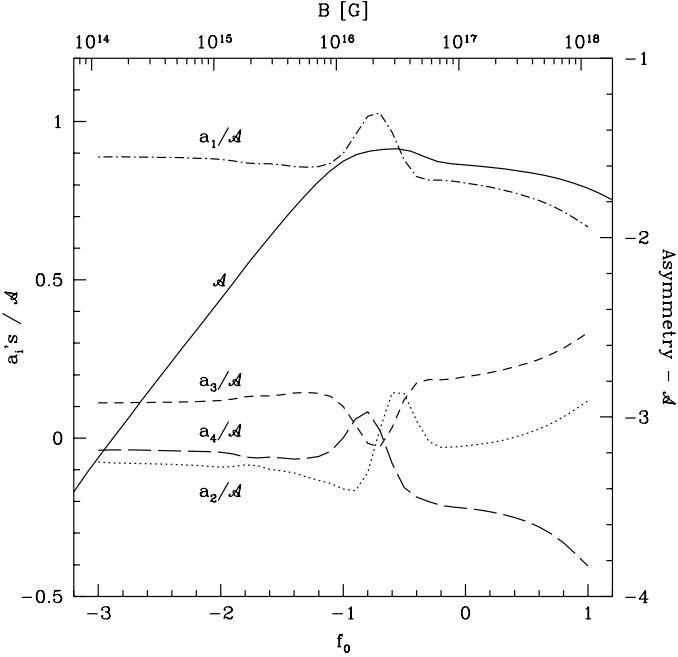,width=\figurewidth}}}
 \caption{The fit parameters and the polar/equatorial asymmetry
 describing the image area for an isothermal NS.}
\label{asym_fit_fig}
\end{figure}}
\figcomment{\figfits}

\subsection{Average properties}
\label{sec:avr}

 When a surface element is emitting black-body radiation into vacuum,
 we know from Stefan-Boltzmann's law that the emitted flux is simply
 $F = \sigma T^4$. Therefore, the intensity of each ray is
 $I=B(T)\equiv\sigma T^4 / \pi$.  This is however changed when the
 element is emitting the black-body photons into a medium with a
 different index of refraction. The reason is that in the presence of
 $n\neq 1$, there are on one hand more photon states for a given
 energy band and on the other, the group velocity which relates the density of
 states to the flux is smaller. For a nondispersive medium with an
 isotropic index of refraction, the two factors are $n^3$ and $1/n$
 respectively, implying that a unit surface will emit a flux of $F=
 \sigma T^4 n^2$, as if the surface area is effectively larger. We can
 therefore define an effective surface area as the equivalent area
 relevant for the Stefan-Boltzmann law of the emitted flux if the
 object would have emitted the photons into vacuum. That is:
\begin{equation}
A_\rmscr{max} \equiv \int {F \over \sigma T^4} dA.
\end{equation}
where $A_\rmscr{max}$ is the effective surface area (and the notation will
soon be clear).

A careful analysis shows however that there there is a second relevant
definition of an effective area. If an objects of surface area $A$ and
temperature $T$ emits radiation into vacuum, then the total luminosity
observed through a distant surface surrounding the $4 \pi$ sterad
around the object, will simply be:
\begin{equation}
L= A \sigma T^4.
\end{equation}
We can therefore define an effective surface area when the
index of refraction is not unity as:
\begin{equation}
\label{eq:Aeff}
A_\rmscr{eff} \equiv {L\over \sigma T^4}.
\end{equation}
Under normal circumstances, we expect that the total flux emitted by a
surface is the total flux observed at infinity. Namely, that
$A_\rmscr{eff}=A_\rmscr{max}$. However, this need not be the case if
some of the light trajectories leaving the surface intersect the
surface again, or in other words, if there is photon
trapping. Generally, we can expect that:
\begin{equation}
A_\rmscr{eff} \leq A_\rmscr{max} (= A n^2~\rmmat{if}~n~\rmmat{is isotropic}).
\end{equation}
There are two effective surface areas and both of them are
interesting. The messy calculation of $A_\rmscr{max}$, the maximum
theoretical effective surface area, is found in the appendix. Its
results are compared with the observed effective surface
area $A_\rmscr{eff}$ calculated here.

The effective area $A_\rmscr{eff}$ of the star which is the {\em
total} area as observered from infinity, is related to the {\em
apparent} area $A'$ seen at various inclinations through the following
average:
\begin{equation}
A_\rmscr{eff} = 4 \int_{i=0}^{\pi/2} A'(i) \cos(i) di,
\end{equation}
The definitions are such that in the unlensed limit, the image has
$A'$ of $\pi R^2$ while the surface areas
$A_\rmscr{max}$ and $A_\rmscr{eff}$ are both $4 \pi R^2$.
The effective surface area can be directly related to the expansion
of $A'$ found in eq.~\ref{fitting_eq} through its integration, giving that
$A_\rmscr{eff}/4 = a_0 + a_1/3 + 7 a_2/15 + 9 a_3/35 + 107 a_4/
315$. It is however calculated directly in order to obtain high
accuracy.

$A_\rmscr{eff}$ is calculated using the ray tracing algorithm. For a
given magnetic field $B_0$ and an observer inclination $i$ above the
magnetic equator, the {\em apparent} surface area $A'$ is calculated.
Since the surface brightness of the isothermal NS is the same in each
direction in which light rays intersect the NS, one can achieve a high
accuracy without the division of the screen into many pixels but
instead use a more sophisticated algorithm. Instead, the accuracy is
achieved through the accurate measurement of the apparent `edge' of
the NS at different angles relative to the center of the screen
(through consecutive bisections) and then an integration over angle
gives an accurate area measurement - $A'(i)$. This area can in turn be
integrated to give $A_\rmscr{eff}$.

The result for different magnetic field strengths can be seen in
figure \ref{figainc} which includes also the total emitted
radiation from the surface $A_\rmscr{max}$ as calculated in the appendix. 

\def\figAeff{
\begin{figure}
 \figcomment{\centerline{\psfig{file=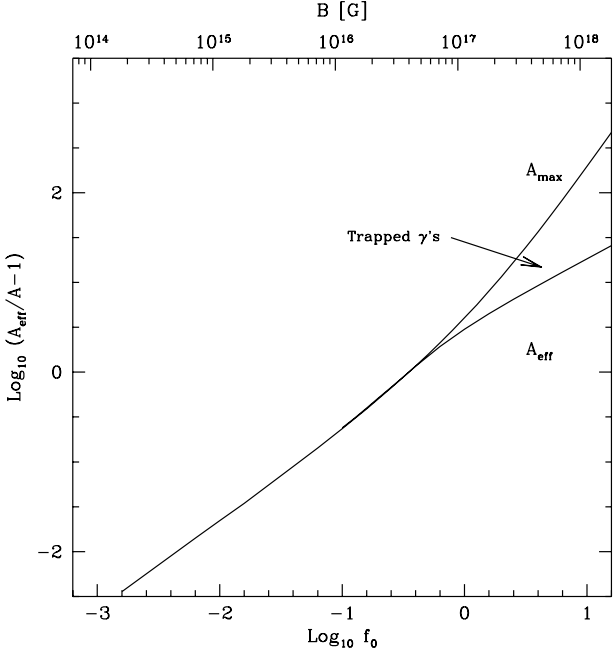,width=\figurewidth}}}
  \caption{The effective area of the NS when observed with the $\perp$
  polarization. For equatorial fields larger than about 
  $5\times 10^{16}$~G,
  the effective area has two definitions: the first
  is defined through the total emitted flux while the second is defined
  as the flux observed at infinity. The former can be larger than the
  latter when some photon trajectories are trapped. For small fields
  one finds a linear relation between the increase in the effective
  area and the field strength.}
\label{figainc}
\end{figure}}
\figcomment{\figAeff}

Clearly from the figure, photons can only be trapped when the
equatorial magnetic field is larger than $2.8\times 10^{16}$~G,
corresponding to the point at which nontrivial distortions of the
apparent images of the NS take place. This also corresponds to the
point where the contribution to $n$ of higher order diagrams than the
one loop approximation may be significant. Hence, this region should
be taken cautiously.

For smaller fields, corresponding to the known field strengths of even
the most highly magnetized PSRs, the apparent surface area in the
$\perp$ polarization is:
\begin{equation}
A_\rmscr{eff}=A_\rmscr{max}\approx A \left( 1+{B\over 5.0\times10^{15}\rmmat{G}} \right ) ~~\rmmat{for}~~ B_0\lesssim  10^{16}~\rmmat{G}.
\end{equation}

\subsection{The light curve}
\label{sec:light_curve}

Generally, the axis of rotation and the magnetic field are not
aligned.  This implies that during the rotation of the star an
observer will view it from a varying magnetic inclination. If the
angular separation between the magnetic axis and the rotational axis
is $\gamma$, then one can relate the magnetic inclination angle $i$ to
that of the inclination above the rotational equator $i_r$ and the
rotational angle $\phi$, which is the rotational phase between the
last time the two axis coincided in the observer's meridional
plane. The relation is:
\begin{equation}
\sin i = \sin i_r \cos \gamma + \cos i_r \sin \gamma  \cos \phi  
\end{equation}
(\cite{Gree83}).  
The best fitting expansion for the apparent area (and therefore the
apparent brightness of an isothermal NS as well) was found to be a
series of $\cos(2 i)$. When expressed directly with $\gamma$, $i_r$
and $\phi$, one then finds that:
\ba
 A'(\phi) \cond &=& a_0 + \sum_{n=1}^4 a_n \left( 1- 2
 \sin^2 i_r \cos^2 \gamma -
\figcomment{\right. \\ \nonumber  & & \left. } 
2 \cos^2 i_r \sin^2 \gamma \cos^2 \phi + \sin(2 i_r) \sin(2 \gamma)
\cos \phi
\right)^n
\label{eq:cos2i}
\ea
Since the odd terms in the expansion dominate the even terms (see
eq.~\ref{eq:expA}), the expansion is predominantly an odd function of
the parenthesized term and the temporal behaviour (i.e., the maxima and
minima) will be determined by this term as well. A simple analysis
shows that for $|i_r|>|\gamma|$, there is one minimum and one maximum
in one rotational cycle. However, for $|i_r|<|\gamma|$, there are two
equal peaks and two different minima in one rotational cycle, as can
be seen in figure
\ref{sample_unpol_fig}. If we neglect the contribution of the even
terms, as we can for the weak field limit, the amplitude of the
variations is just:
\begin{equation}
A'_\rmscr{max} - A'_\rmscr{min}  =  a_1 F_1(\gamma,i_r) + a_3 F_3(\gamma,i_r),
\end{equation}
where we have defined two auxiliary functions as:
\begin{eqnarray}
F_1 (\gamma,i_r)  \cond &=& \cond \left\{
\begin{array}{lr}
2 \sin (2 i_r) \sin (2 \gamma), & ~~~~ |i_r|>|\gamma|  \\
2 \sin^2(i_r + \gamma), & ~~~~~ |i_r|<|\gamma| \end{array} \right. 
\end{eqnarray}
\begin{eqnarray}
F_3 (\gamma,i_r) \cond &=& \cond \left\{
\begin{array}{l}
 {1\over 2} \left(3 \sin(2
 i_r) \sin(2 \gamma) 
\figcomment{ \right. \\ \hfill\left. } 
+ \sin(6 i_r) \sin(6 \gamma) \right), 
 ~ \hfill |i_r|>|\gamma| \nonumber \\
 \left(3+2 \cos(2(ir+\gamma)) 
\figcomment{ \right. \\ \hfill \left.}
+\cos(4(i_r+\gamma)) \right)
 \sin^2(ir+\gamma),
  ~ \hfill |i_r|<|\gamma| \end{array} \right.
\end{eqnarray}
%
The two functions incorporate the two regimes for which there are
either one minimum or two minima in one rotation period.  The
difference between the depth of the two minima in the second case (when 
$|i_r|<|\gamma|$) is the amplitude found if the $|i_r|>|\gamma|$ regime
is used for the evaluation of $F_1$ and $F_3$.
 If $a_1$ is
negative, the peaks become valleys and vice versa.

A rotating isothermal neutron star observed in unpolarized light
will therefore have light variation of which the amplitude in the weak
field limit is given by,
\begin{eqnarray}
 {\Delta L \over L} & =& {A'_\rmscr{max}-A'_\rmscr{min} \over A +
 A'_\rmscr{avr}} 
\figcomment{ \\ \nonumber &}
\approx 
\figcomment{ & } 
\left( F_1(\gamma,i_r) + 0.12 F_3(\gamma,i_r)
 \right) \left( B \over 6.3\times10^{17}~\rmmat{G} \right).
\end{eqnarray}

The functions $F_1$ and $F_3$ are linear for small $\gamma$'s and
linear in the colatitude $\pi/2-i_r$. For general values of $\gamma$
and $i_r$, they are of order unity. Namely, for small magnetic fields,
the effect will fall as $\propto \gamma B^2$ when viewed at random
viewing angles and small separations between the axes. 
A few sample light curves are seen in figure \ref{sample_unpol_fig}.
\def\figunpol{
\begin{figure}
\figcomment{\centerline{\psfig{file=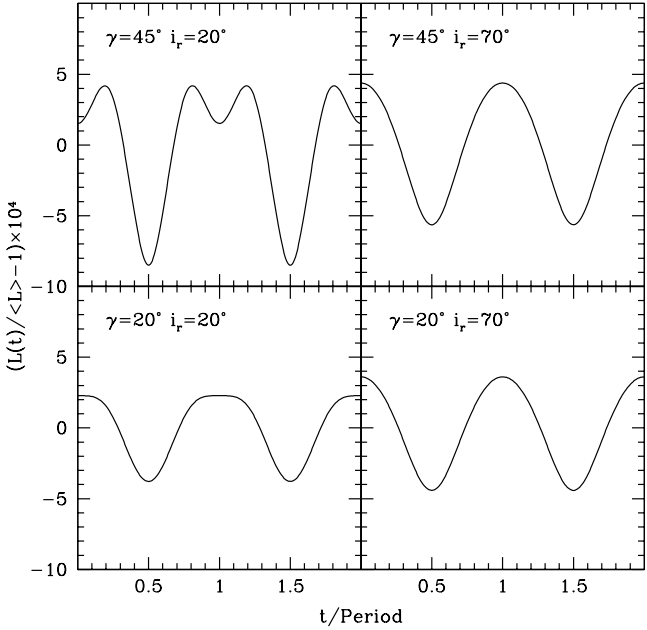,width=\figurewidth}}}
 \caption{ A few sample light-curves of an isothermal NS having an
 equatorial field of $10^{15}$~G, observed in unpolarized light, for
 different magnetic axis - rotational axis separations $\gamma$ and
 different rotational inclination angles $i_r$.  }
\label{sample_unpol_fig}
\end{figure}}
\figcomment{\figunpol}

A more interesting effect however arises when studying the polarized
light curves. In the unpolarized case, the effect stems from the
asymmetry between the polar image and the equatorial image, an
asymmetry that is never large, even for fields where $n-1$ approaches
unity. If we, however, look in polarized light, we have two images of
the neutron star of which the area difference can be large for
intermediate fields. Moreover, we shall soon see that irrespective of
$\gamma$, the effect is always large and falls only as $\sim B$ for
small fields (as long as $B\gtrsim B_\rmscr{QED}$). Thus, it is
expected to be more prominent.

Using the polarization locking behaviour of the light rays (see
\S{\ref{sec:polar}), we know that photons leaving the surface with a
$\para$ or $\perp$ polarization will keep their polarization until a
large distance from the star at which point the local magnetic field
lines can be assumed to be totally aligned with the apparent direction
of the magnetic axis (with the direction of the axis' projection onto
the observer's plane of the sky). Since the magnetic axis is rotating,
the two primary polarization directions are the projected
direction of the axis onto the plane of the sky and its perpendicular
direction within this plane.

If we work in a coordinate system aligned with the rotational
$z$-axis, and a $y$-axis that is perpendicular to the plane containing
the line of sight and the $z$-axis, then the observer's direction is:
\def\obs{{\hat{\bf o}}}
\def\mag{{\hat{\bf m}}}
\begin{equation}
\obs=\cos i_r \, \xd + \sin i_r \, \zd
\end{equation}
If we use the rotational phase $\phi$ and the separation $\gamma$
between the two axes, the direction of the magnetic axis is:
\begin{equation}
\mag = \sin \gamma \cos \phi \, \xd + \sin \gamma \sin \phi \, \yd 
+ \cos \gamma \, \zd
\end{equation}
Using this, we can calculate the cosine of the {\em apparent} angle
$\varpi$ that the magnetic axis makes with the $y$-axis (it is also
the horizontal direction of the observer's screen). To do so, we
project the magnetic axis to the plane of the sky and calculate its 
dot product with the observer's $y$-axis (which is
also our coordinate system $y$-axis):
\begin{eqnarray}
 \cos^2 \varpi \cond &=& \cond {\left((\mag - (\mag\cdot \obs) \obs) \cdot
 \yd\right)^2 \over \left|\mag - (\mag\cdot \obs) \obs\right|^2} 
\figcomment{ \cond \\ \nonumber & }  
= 
\figcomment{ & \cond }
 {\left( \sin\gamma \sin\phi \right)^2 \over 1-(\cos \gamma \sin i_r +
 \cos i_r \sin \gamma \cos \phi)^2 } \equiv p(\gamma, i_r, \phi)
\end{eqnarray}
If we neglect the slight asymmetry between the apparent size when
observed from the poles or from the equator, then the images observed
at one or the other polarizations that are aligned with the magnetic
axis are going to be fixed at $A$ and $A_\rmscr{eff}$. If however our two
observational polarization directions are fixed at for example the
rotation axis and the perpendicular direction, then the polarized
light-curve observed will vary because of the varying component
observed from each polarization.

The brightnesses of the two polarizations fixed to the magnetic axis are
\begin{eqnarray}
 L_\para & = &{A \over A + A_\rmscr{eff}} L \approx {1\over 2} \left( 1 -
 {\Delta A \over A}\right) L ~~\rmmat{for}~~\Delta A<< A \\ L_\perp & =
 &{A_\rmscr{eff} \over A + A_\rmscr{eff}} L  \approx {1\over 2} \left( 1+ {\Delta A
 \over A}\right) L ~~\rmmat{for}~~\Delta A<< A
\end{eqnarray}
The two light curves at the two polarizations will therefore be:

\begin{eqnarray}
 L_{1} & = & L_\perp \cos^2 \varpi + L_\para \sin^2 \varpi 
\figcomment{ \\ \nonumber &}
\approx
\figcomment{&}
 {L\over 2} + {L\over 2}{\Delta A \over A} \left(2 p(\gamma, i_r,
 \phi)-1\right). \\ L_{2} & = & L_\perp \sin^2 \varpi + L_\para \cos^2
 \varpi 
\figcomment{ \\ \nonumber &}
\approx
\figcomment{&}
{L\over 2} - {L\over 2}{\Delta A \over A} \left(2
 p(\gamma, i_r, \phi)-1\right).
\label{eq:two_pols}
\end{eqnarray}
where for small fields one has
\begin{equation}
 {\Delta A\over A} \approx  {B \over 5.0\times 10^{16} G}
\end{equation}

This light curve normally has two peaks during one rotational cycle
but the peaks do not have to be symmetric. The averages of each
polarized light curve need not be equal, as can be seen in figure
\ref{sample_pol_fig}.  However, irrespective of whether there is an
asymmetry and irrespective of $\gamma$, one normally expects to see
two different average amplitudes for the two different polarizations.

Another interesting point is that even if the magnetic and rotational
axes are aligned and there is no variation at all of the light curves
during rotation, one still finds that the two polarizations will have
a different amplitude:
\begin{equation}
{\Delta L \over L} = {L_2 - L_1 \over L} = {L_\perp - L_\para \over L}
= {\Delta A \over A}.
\end{equation}

\def\figpol{
\begin{figure}
\figcomment{\centerline{\psfig{file=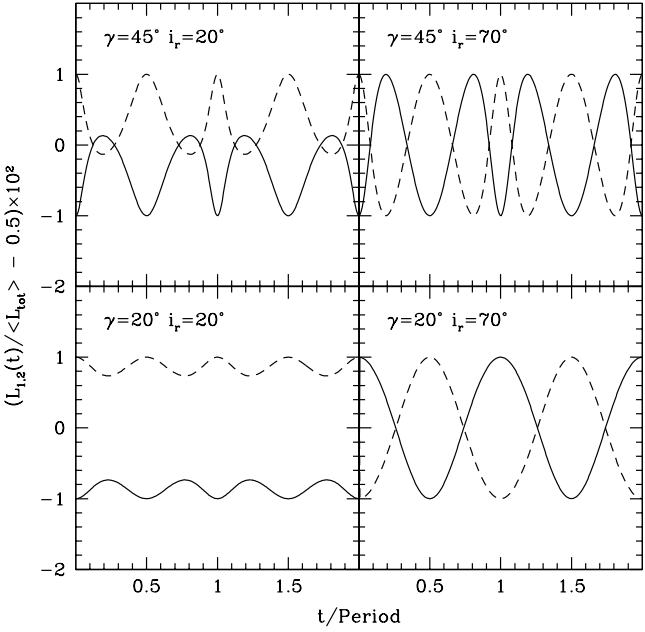,width=\figurewidth}}}
 \caption{ A few sample light-curves of an isothermal NS having an
 equatorial field of $10^{15}$~G observed in polarized light, for
 different separations ($\gamma$) between the magnetic and rotational
 axes and different rotational inclination angles $i_r$. The two
 curves are the observed light curve in the two polarizations parallel
 and perpendicular to the rotational axis. (Solid line - when the
 ray's $B$ field is parallel to the rotational axis, dashed - when it
 is perpendicular to the axis.}
\label{sample_pol_fig}
\end{figure}}
\figcomment{\figpol}

\section{A realistic temperature distribution}
\label{sec:realistic}

Neutron stars are not in thermal equilibrium.  They are born with
$T\gtrsim 10^9$~K (\cite{Shap83}).  After about one hundred years,
their cores become isothermal and cooling proceeds quasistatically.
Until $10^5 - 10^6$~yr have passed, the neutrinos dominate the heat
flux from the core (\cite{Shap83}, \cite{Heyl98numens}).  Later photon
emission from the surface carries heat from the neutron star.
Regardless of the cooling process dominating at a particular time, the
energy to supply cooling emission from the surface must traverse the
thin insulating layer surrounding the neutron star core, the envelope.
For warm and hot neutron stars the most important vehicle for the heat
flow is electron conduction (\cite{Gudm82}).

A strong magnetic field dramatically alters the conduction properties
of neutron star envelopes (\cite{Hern84a}).  Electrons flow more
easily along the field lines than across them, which causes the flux
distribution across the surface to be anisotropic (\eg \cite{Scha90a},
\cite{Scha90b}, \cite{Heyl97analns}, \cite{Heyl98numens}).  For
magnetic fields stronger than $10^{14}$~G, analytic techniques can model
the thermal structure of the envelope with adequate precision.
\jcite{Heyl97analns} find that a simple prescription describes the
crust flux distribution for these highly magnetized neutron stars:
\be
F_c \propto B^{0.4} ( {\usrf} \cdot {\hat{\bf B}} )^2,
\label{eq:real_flux}
\ee
where $\usrf$ is a unit vector normal to surface, and $B$ and $\hat{\bf B}$
are the magnitude and direction of the magnetic field.

This indicates that in addition to the effect of the asymmetry between
the polar image and the equatorial image that arises from the magnetic
lensing, a large asymmetric contribution will arise from the flux
gradient along the surface. Hence, it is very likely that the latter
asymmetry will drown the unpolarized effect of \S\ref{sec:light_curve}.

When calculating the emitted fluxes, one has to take into account that
it is the flux driven through the crust which is fixed constant and
not the surface temperature. In other words, a surface element will emit a
total flux given by eq.~\ref{eq:real_flux}, even if it is a more
efficient black-body radiator.

Due to the increased number of states in one polarization (per unit
energy) and the reduced group velocity, the black-body radiation
law becomes:
\begin{equation}
 F_\perp = \sigma T^4 \left( 1 + g(B,\psi) \right)/2~~{\rm
 and}~~ F_\para = \sigma T^4/2,
\end{equation}
where the function $g(B,\psi)$ describes the increase in flux due to
the magnetic field strength $B$ that makes an angle $\psi$ with the
surface normal. A fitting function for it in the limit of weak fields
is found in the appendix.

If the equilibrium temperature without the magnetic field would have
been $T_0$, then it is evident that the new equilibrium temperature is
going to be smaller in order to keep the same outgoing flux constant:
\begin{equation}
T^4 = {T_0^4 \over 1+g(B,\psi)/2}.
\end{equation}

When we now ray-trace to find the resulting image, we have to take
into account that the flux driven into the surface is a function of
the magnetic field and that the temperature is a function of the field
as well. Each ray coming from the observer's screen that intersects
the surface will have, irrespective of the polarization, a radiation
intensity of:
\begin{equation}
I(B,\psi) = 
{1\over 2}\left(B \over B_\rmscr{pole} \right)^{0.4} {  \cos^2 \psi \over
1+g(B,\psi)/2} {1\over \pi} \sigma T_\rmscr{pole}^4,
\end{equation}
where $T_\rmscr{pole}$ is the polar effective black-body temperature.
Excluding the index of refraction effect that enters through the
denominator, the flux emerging from the poles will be larger than that
of the equator. As a consequence, the image observed from the polar
direction will be brighter than the image from equatorial
inclinations. The effect of the index of refraction is to slightly
cool the star.  Since the image in the $\para$-mode is unaffected by
the nontrivial $n$, it will be fainter. On the other hand, the
increased image surface area of the $\perp$-mode will imply that it is
brighter than the unlensed image, even though the lensed star is
cooler. If averaged over the $4\pi$ sterad around the star, the sum of
both images should yield the same total flux that it would have had
without including of the refraction effects.

To find the light curves, we find fitting formulae to the total image
brightness as a function of magnetic inclination. They are accurate to
roughly 1 part in $10^3$ for very small fields.  First, the we fit
image brightness in the limit of $B\rightarrow 0$. It is:
\begin{equation}
 L_{\para}(B\rightarrow 0) = L_{\perp}(B\rightarrow 0) 
 = 
L_0 \left(
 b_0 + \sum_{n=1}^4 b_n \cos^n (2i) \right) \nonumber 
\end{equation}
\[
 b_0  =  0.5205,~~b_1=-0.1318,~~b_2=0.011,~~
\figcomment{ \] \[ }
b_3=-0.0026,~~{\rm
 and~~} b_4=0.0007 .
\]
$L_0\propto \sigma T_\rmscr{pole}^4$ is the brightness that would have been
observed if the NS would have had no magnetic field and would have had
an isothermal temperature distribution with $T_\rmscr{pole}$.

Next, we fit for the corrections to the observed brightnesses for each
of the two polarizations:
\begin{equation}
 L_{\para} = L_{\para}(B\rightarrow 0) + L_0 \left( b_{\para,0} +
 \sum_{n=1}^4 b_{\para,n} \cos^n (2i) \right) {B \over B_0} 
 \nonumber
\label{eq:noncf_lpara}
\end{equation}
\[
b_{\para,0} =-0.7792,~~b_{\para,1} =
 0.2546,~~b_{\para,2} = -0.0077,~~
\figcomment{\] \[}
b_{\para,3} = 0.0001,~~
 \rmmat{and}~~
 b_{\para,4}=0.0048 ,
\]
\begin{equation}
 L_{\perp} =
 L_{\perp}(B\rightarrow 0) + L_0 \left( b_{\perp,0} + \sum_{n=1}^4
 b_{\perp,n} \cos^n(2i) \right) {B\over B_0} \nonumber 
\label{eq:noncf_lperp}
\end{equation}
\[ 
b_{\perp,0} =0.4069,~~b_{\perp,1} = 0.4960,~~b_{\perp,2} =
 -0.0753,~~
\figcomment{\] \[}
b_{\perp,3} = -0.0429,~~\rmmat{and}~~ b_{\perp,4}=0.1056 .
\]

With these fits, we can construct any light-curve in the limit of small
magnetic fields. The polarization brightnesses vary by more than 50\%
when observed at different latitudes, however, if we look at the ratio
between the two, we single out the effect of the magnetic
refraction. The ratio (for small magnetic fields) normalized to a
magnetic field strength of $B_0$ is depicted in figure
\ref{pol_ratio_fig}. For $10^{15}$~G, one finds that the ratio varies
between roughly $1.015$ to $1.03$ when the magnetic inclination is
varied from the pole to the equator. The reason that the equatorial
image brightness ratio is larger than the polar one is because the
``over-the-horizon'' effect of the magnetic refraction affects the
equatorial image more because the luminous poles lie on the horizon --
its inclusion is more important than the inclusion of the faint
equatorial regions in the polar images.

\def\figpolratio{
\begin{figure}
\figcomment{\centerline{\psfig{file=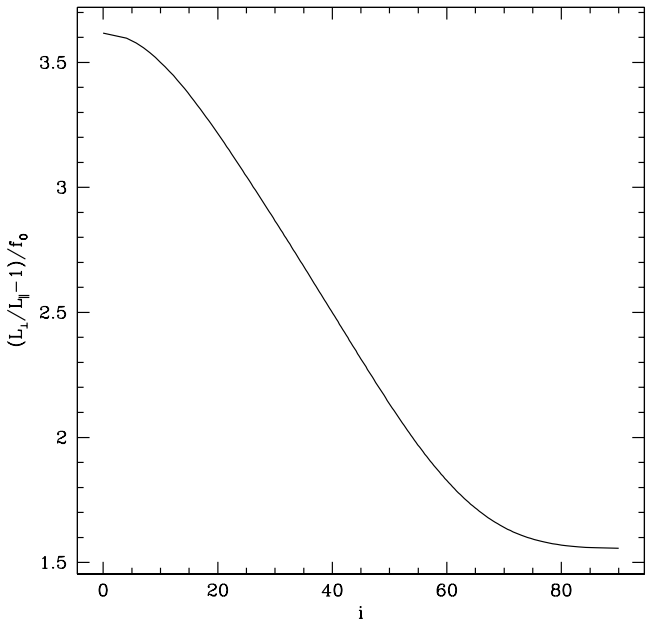,width=\figurewidth}}}
 \caption{The weak field ratio between the total flux observed in two
 polarizations as a function of the observer's magnetic inclination
 $i$, normalized to $f_0$. For small magnetic fields, the ratio
 $L_\perp/L_\para$ is $1$ plus the graphed function times $f_0$. }
\label{pol_ratio_fig}
\end{figure}}
\figcomment{\figpolratio}

To construct the light-curves as a function of $i_r$ - the inclination
above the rotational axis, $\gamma$ - the separation angle between the
two axes, and $\phi$ - the rotational phase of the star, one first
needs to calculate for given angles the expression for $\cos(2i)$, as
is given in eq.~$\ref{eq:cos2i}$. This allows the calculation of
$L_\para$ and $L_\perp$ through the use of
eqs.~\ref{eq:noncf_lpara}~and~\ref{eq:noncf_lperp}.  Then, using
eq.~\ref{eq:two_pols} one can obtain the polarized light curves when
observed in the two directions that are aligned with and perpendicular
to the rotational axis of the star. A few sample polarization ratio
light-curves are found in figure \ref{non_conflux_lc}. One set of
curves describes the {\em total} polarization, \ie, when the
measurement directions follow the $\para$ and $\perp$ polarizations,
while the second set of curves is a more realistic case in which the
observer's polarization planes are fixed and aligned with or
perpendicular to the rotational axis.

\def\fignoncf{
\begin{figure}
\figcomment{\centerline{\psfig{file=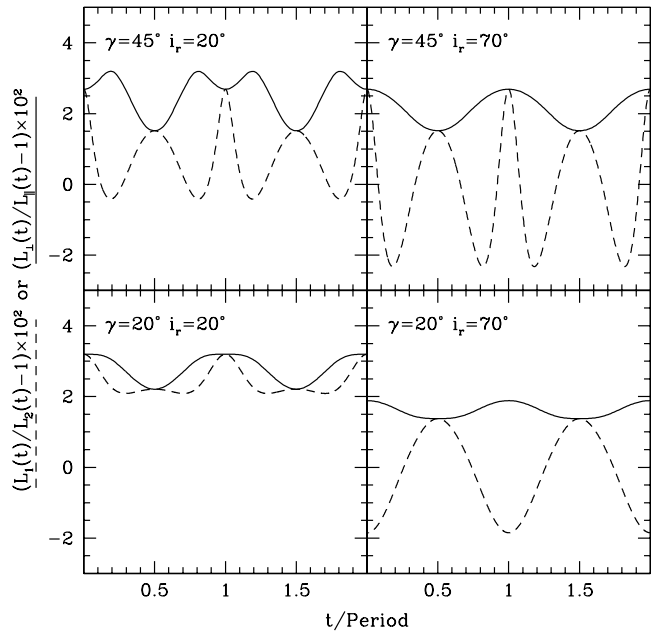,width=\figurewidth}}}
 \caption{Several sample polarizations light curves predicted for a
 $10^{15}$~G neutron star. For different viewing inclinations $i_r$
 above the rotation equator one will observe a changing ratio between
 the two polarizations. The ratio $L_{\perp}/L_{\para}$ (solid line)
 is the flux ratio of the two polarizations while having the
 observer's polarization planes follows the two polarizations
 states. The $L_1/L_2$ ratio (dashed line) is the ratio predicted if
 the observer's polarization directions are aligned with the
 rotational axis of the neutron star.}
\label{non_conflux_lc}
\end{figure}}
\figcomment{\fignoncf}


\section{Discussion}
\label{sec:discuss}

The magnetic field surrounding a neutron star magnifies and distorts
the image of its surface.  If the surface field is everywhere less
than $3\pi/\alpha \times B_\rmscr{QED} \approx 5.7 \times 10^{16}$~G, the
distortion is trivial; the neutron star appears slightly larger, and
some light from the far side of the star contributes (the left side of
figure~\ref{sample_images}).  In this regime, only the image in the
perpendicular polarization is magnified.  If each element of the
neutron star's surface radiates as a blackbody, it will appear 3.2 \%
brighter in the perpendicular polarization than in the parallel
polarization for the inferred surface field of SGR~1806-20.  The flux
difference between the two polarizations increases linearly with the
strength of the field at the surface.

Furthermore, as the neutron star rotates, the image in a fixed
polarization consists of a varying admixture of the images of the
parallel and perpendicularly polarized photons.  The phase of this
oscillation at a given time depends weakly on the energy of the
photons.  The observed polarizations of higher energy photons reflect
the polarization modes of the neutron star slightly later in its
rotation period (eq.~\ref{eq:decouple}).  

In regions of the magnetosphere where the field strength exceeds $5.7
\times 10^{16}$~G, the index of refraction for perpendicularly
polarized photons will exceed $3/2$, and the relationship between the
group and phase velocities of the light rays becomes nontrivial.
Regions of the star appear in the image several times with differing
parities.  The index of refraction is neither isotropic with respect
to the direction of the photon's propagation nor isotropic around the
star; therefore, the optically dense magnetosphere produces a
complicated image of the neutron star surface (the right side of
figure~\ref{sample_images}).

Although the full complexity of the images created by supercritical
magnetic fields surrounding neutron stars will be difficult to observe
in the near future, several hallmarks of these effects can be observed
with some of the next generation of X-ray telescopes (e.g.,
Spectrum-X-Gamma), if sufficient precision can be achieved to observe
the effects averaged over the two polarizations.  If a particular type
of emission is restricted to small regions of the magnetosphere of the
neutron star (\eg curvature and synchrotron emission may be restricted
to the polar regions), this emission may be strongly magnified even in
weak fields. Figure~\ref{sample_images} depicts a neutron star whose
apparent area is about 50\% larger than its actual area.  The areas
above 80$^\circ$ latitude are magnified by factor of up to 2.3
depending on inclination.  Generically the polar regions will be
magnified twice as much as the star on average since the polar field
is twice as large.  In the strong field limit, the complex topology of
the images would be reflected in a complicated light curve for
emission localized to the polar regions.  Several pulses with varying
parities could be visible from each polar region as the star rotates.

Time-resolved spectropolarimetry of strongly magnetized neutron
stars would reap substantial rewards.  All of the effects discussed in
this paper affect only the perpendicular polarization mode.  High
energy photons traveling in the parallel mode will suffer photon
splitting and transfer their energy into the perpendicular mode
(\cite{Adle71}, \cite{Bari95a}, \cite{Heyl98cascade}).  This second
process will affect the spectra of these sources especially for 
$E \sim m_e c^2$.  The lensing process discussed here affects even low
energy photons ($E \ll m_e c^2$) -- the effect will be even stronger
at higher energies since the index of refraction increases with photon
energy (\cite{Adle71}).  Future generations of X-ray telescopes may
be able to measure not only the energy but also the polarization of the
photons that they detect (R. Rutledge, private communication).  

Even in weak fields, the perpendicular polarization from the polar
regions would subtend a larger solid angle in the image than the
parallel polarization; consequently, its pulse would last longer.
Moreover, the gross magnifications of the star and the augmented
magnification of the polar regions would change the light curves in
the perpendicular polarization both in magnitude and character.  For a
uniform temperature blackbody, the perpendicular polarization would be
brighter than the parallel polarization and exhibit slight
variability.  The parallel polarization would be what one would expect
for a blackbody at the appropriate temperature, modulo the effects of
photon splitting.  If the star has a nonuniform temperature
distribution with the poles hotter than the equatorial regions, the
lensing will increase the variability observed in the perpendicular
polarization, and also increase the total flux detected in that
mode.

Several additional effects could quantitatively change the results,
two of which are gravitational lensing and the presence of higher
multipole fields. Although gravitational lensing does not compete
directly with magnetic lensing in the sense that its effect is
independent of polarization (and thus creates no polarization
signal), it can slightly smear the signal by adding over-the-horizon
contributions.

The effects of higher multipoles can on the other hand be potentially
more important. Multipoles that are somewhat higher than the dipole
will be directly observable as they will contribute higher Fourier
components to the polarization light curve (e.g.~fig.~8). Under
favourable conditions, it could allow the measurement of each
multipole separately. Very high multipoles will not have a varying
signal since the observation of a whole hemisphere will average their
variations. Nevertheless, since an average field strength does exist,
a constant polarization signal of $L_\perp/L_\para \sim 1+ min\{
B_{multipole}/10^{17}G, \Delta R/R\}$ will be introduced.
$\Delta R$ is the typical height above the surface on which the high
multipole field decays. It bounds the maximal $L_\perp$ since
the increase of the apparent area cannot be larger than the
cross-section of the region where the index of refraction is not
unity. Although multipoles can complicate the analysis, they could in
principle be detected if they were more important than the dipole
field.

\section{Conclusions}
\label{sec:conclus}

We have studied how photons travel through the magnetospheres of
strongly magnetized neutron stars.  For surface fields less than 
$3\pi/\alpha \times B_\rmscr{QED} \approx 5.7 \times 10^{16}$~G, the 
magnetosphere distorts the image of the neutron star surface
minimally.  The total area of the image in the perpendicular
polarization is magnified by $B/(5.0 \times 10^{14}~\rmmat{G})$
percent, and some light from the far side of the star contributes.
For fields stronger than $5 \times 10^{16}$~G, the distortion of the
image becomes more complicated.  Regions of the stellar surface appear
several times with varying parity in the image.  

Unlike gravitational lensing of neutron stars (\cite{Page95},
\cite{Heyl97analns}), magnetic lensing only affects the perpendicular
mode and tends to increase the variability of the source;
consequently, by measuring how the polarized flux varies as the
magnetar rotates, one could observe the lensing described in this
paper and estimate the strength and inclination of the surface
magnetic field in a manner independent of the details of the dipolar
spin down.

\section*{Appendix A: The maximal effective surface area of a magnetar}

In \S\ref{sec:avr} we have seen that there are two definitions for the
effective area of a magnetar -- the area required to emit the same
black body luminosity by an object without a magnetic field.  The
first is the effective area to an observer at infinity (using the
total luminosity observed there) while the other is the effective area
to the object itself (using the total emitted luminosity). The
definitions are not the same because some of the emitted photons can
be reabsorbed by the star through trapped photon trajectories.

In \S\ref{sec:avr} we have calculated the effective area seen at
infinity. To see whether photon trapping exists, one has to calculate
the total luminosity emitted by the star.

The effective area increase of a surface element is given by the
relative increase in the black-body flux. Since the index of
refraction $n$ is not a function of energy, the flux increase is
simply an angular integration:

\begin{equation}
 {da' \over da} = {f' \over f} = {1\over \pi} \int_{\vg(\kv)\cdot\usrf
 >0 } d \Omega_{k} n(\kv)^3 \left( \vg \cdot \usrf \right)
\label{flux_inc_eq}
\end{equation}

The integral is over the half sphere for which the group velocity
$\vg$ points upwards, away from the surface (of which the normal is
$\usrf$).  The $n(\kv)^3$ factor is the increase of the number of
possible photon states for a given energy interval at a direction
$\Omega_k$. The factor $\vg(\kv)\cdot\usrf$ is the contribution to the flux
perpendicular to the surface from these states. For an isotropic
nondispersive medium, one immediately finds, as one should, that
$f'/f=n^2$.

The integral boundaries might appear simple but they are in fact very
complicated because $\vg$ is generally not in the direction of
$\kv$. We can however simplify the boundaries through the use of a
symmetry property of $n(\kv)$. Since $n$ is parity conserving:
$n(-\kv)=n(\kv)$ and since $\vg$ is a vector that is related to $n$
through a gradient, it will satisfy: $\vg(-\kv)=-\vg(\kv)$. If we now
observe figure~\ref{fig:integration} describing the vertical slice of
the integration regions through the plane containing ${\bf B}$ and
$\usrf$, then it is clear that each point within the proper
integration region (e.g. point 1 or 2) has an equivalent counterpart
with an opposite group velocity outside the integration region
(e.g. point 1' or 2'). Thus, to get the required integral, we can
either integrate using the cumbersome boundaries or integrate the
absolute value of the integrand of the whole region and divide by
2. We choose to do the latter in the polar coordinates
$(r,\theta,\phi)$ aligned with ${\bf B}$ such that the $\hat{\bf y}$
axis is perpendicular to the plane containing both ${\bf B}$ and
$\usrf$.

\def\figintegration{
\begin{figure}
\figcomment{
\centerline{\psfig{file=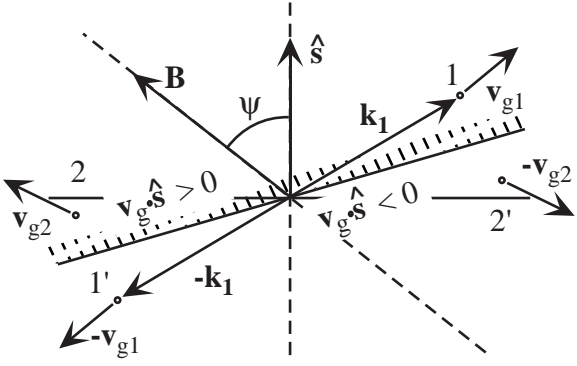}}
}
 \caption{The $k$-space over which the flux integration should be
 carried is the region in which ${\bf v}_g \cdot \usrf > 0 $ -- \ie,
 above the hatched line. Each point in the integration region
 (e.g. 1,2) has however a counterpart outside the region (1',2') for
 which the group velocity is opposite. If we integrate $|{\bf v}_g
 \cdot \usrf|$ over all space we get twice the required answer without
 the complication arising from the boundary.}
\label{fig:integration}
\end{figure}}
\figcomment{\figintegration}

In our coordinate system, we have that the radial unit vector is:
\begin{equation}
 \hat{\bf r} = \sin \theta \cos \phi \,\hat{\bf x} +
\sin \theta \sin \phi \,\hat{\bf y} + \cos \theta \,\hat{\bf z}.
\end{equation}
The unit vector in the direction of $\theta$ is needed for the group
velocity, it is:
\begin{equation}
 \hat{\bf \theta} = \cos \theta \cos{\phi} \,\hat{\bf x} +
\cos \theta \sin \phi \,\hat{\bf y} - \sin \theta \,\hat{\bf z}.
\end{equation}
Last, we need the vector in the direction normal to the surface
element. If the angle between the magnetic field direction and the
normal to the surface is $\psi$, then this vector is:
\begin{equation}
\usrf= -\sin \psi \,\hat{\bf x} + \cos \psi \,\hat{\bf z}.
\end{equation}
The group velocity is related to $\omega$ through:
\begin{equation}
 \vg = \nabla_{k} \omega = \left( \partial \omega \over \partial k
 \right)_\theta \,\hat{\bf r} - {1\over k} \left( \partial \omega \over
 \partial \theta \right)_{k} \,\hat{\bf \theta}.
\end{equation}
If the index of refraction is in the form $n= 1+f(B) \sin^2 \theta$,
then we find:
\begin{equation}
 \vg = {1 \over 1+ f(B) \sin^2 \theta} \,\hat{\bf r} - {2 f(B) \sin
 \theta \cos \theta \over (1+ f(B) \sin^2 \theta)^2} \,\hat{\bf \theta}.
\end{equation}
The integrand is therefore:
\begin{eqnarray}
 n^3 \left| \vg \cdot \usrf \right| & = &\left| \left(1+f(B) \sin^2
 \theta  \right)^2  
\figcomment{ \times \right. \nonumber \\ & & \left. }
\left( \cos \psi \cos \theta - \sin \psi \sin
 \theta \cos \phi\right) \right. \nonumber \\ & & \left. + 2 f(B) \sin
 \theta \cos \theta \left( 1+ f(B) \sin^2 \theta \right) 
\figcomment{ \times \right. \nonumber \\ & & \left.}
\left( \sin \psi
 \cos \theta \cos \phi + \sin \theta \cos
 \psi\right)^{\phantom{1}}\hskip -4pt  \right|.
\end{eqnarray}

Next, we need to integrate over the NS surface with its varying field
strength and varying directions. Since we assume it to be a dipole, we
have that if the equatorial magnitude of the magnetic field is
$B_\rmscr{eq}$ then on the surface we have:
\def\hn{\usrf}
\def\hm{\hat{\bf m}}
\begin{equation}
{\bf B} = B_\rmscr{eq} \left( 3 (\hn \cdot \hm) \hn  - \hm \right).
\end{equation}
 We thus find that the magnitude of ${\bf B}$ and its angle from the
 local normal are:
\begin{eqnarray}
 B & = &\sqrt{3 (\hn \cdot \hm)^2 +1 } B_\rmscr{eq} = \sqrt{3 \cos^2 \alpha
 +1 } B_\rmscr{eq} \\ \cos \psi & = & \hat{\bf B} \cdot \hn = {2 \cos \alpha
 \over \sqrt{3 \cos^2 \alpha +1}}.
\end{eqnarray}
We further use that $f(B)$ is linear in $B$ in the interesting regime,
implying that $f(B) = f(B_\rmscr{eq}) B \equiv f_0 \sqrt{3 \cos^2 \alpha +1
}$.

The growth of the effective area of the whole star can now be written as:
\begin{equation}
 {a' \over a} = {1\over a} \int da' = {1\over a} \int da {da' \over
 da} = {1\over 4 \pi} \int d\Omega_a {da' \over da},
\end{equation}
where $\int d\Omega_a$ implies an integration over the surface
colatitude and longitude $(\alpha,\varphi)$. The integration over
$\phi$ is trivial and we have:
\begin{eqnarray}
 {a' \over a} \hskip -3pt &=& \hskip -3pt 
{1\over 4 \pi} \int_{1}^{-1} d(\cos \alpha)
 \int_{1}^{-1} d(\cos \theta) \int_0^{2 \pi} d\phi 
\figcomment{ \times \nonumber \\ & & }
\left| \left(1+f(B)
 \sin^2 \theta \right)^2 \left( \cos \psi \cos \theta - \sin \psi \sin
 \theta \cos \phi\right) \right. \nonumber \\ & & \left. + 2 f(B) \sin
 \theta \cos \theta \left( 1+ f(B) \sin^2 \theta \right) 
\figcomment{ \right. \times \nonumber \\ & & \left. }
\left( \sin
 \psi \cos \theta \cos \phi + \sin \theta \cos
 \psi\right)^{\phantom{1}}\hskip -4pt \right|,
\end{eqnarray}
where we explicitly have to express $f(B)$ as $f_0 \sqrt{3 \cos^2
\alpha +1}$ and $\cos \psi$ as $2 \cos \alpha / \sqrt{3 \cos^2 \alpha
+1}$. The integration is nontrivial (mainly because of the absolute
sign) but is however straight forward if done numerically. The method
used was a Monte Carlo integration which easily achieved accuracies
better than $10^{-3}$ for several dozen magnetic field strengths in
roughly an hour of an HP workstation CPU time.

Another useful computation is the direct evaluation and the
construction of a fitting formula for the local increase in the
black-body flux as given by eq.~$\ref{flux_inc_eq}$. The integral is
generally a nonlinear function of both the magnetic field strength $B$
and the angle $\psi$ between the magnetic field, ${\hat {\bf B}}$, and
the surface normal, $\usrf$. Thus, a simple fit can only be found in
the limit of weak fields ($B \ll B_0 \approx 1.13\times10^{17}$~G), it
is:
\begin{eqnarray}
1+ g(B,\psi) \cond &\equiv& \cond 
{da' \over da} \approx 1+ \left( 1+0.067 \cos \psi +
1.6024 \cos^2 \psi 
\figcomment{ \nonumber \right. \\ & & \left.}
-0.672 \cos^3 \psi \right) \left( B / B_0\right)
\end{eqnarray}
Its relative accuracy is better than $10^{-3}$, which is the
integration accuracy.

\figuncomment{\clearpage}


\figcomment{
\end{document}